# Comparison of anomaly detection and supervised learning for acute myeloid leukemia blast identification: A proof-of-concept study

Mezya Sezen[1,*], Pavinap Priyaa[1,*], Krishnapriya Vinod[1], Atul Thatai[2], Nitin Dayal[2], Ragothaman M. Yennamalli[3$], Rama S. Akondy[1$].

[1]Department of Biology, Trivedi School of Biosciences, Ashoka University, Sonipat, India 131029

[2]Max Healthcare Institute Ltd., New Delhi, India, 110017

[3]School of Computational and Integrative Sciences, Jawaharlal Nehru University, New Delhi, India 110067.

[*] These authors contributed equally to this manuscript.

[$] Corresponding Authors

## Abstract

**Background.** Flow cytometry is the established standard for identifying leukemic blast populations in AML, but relies predominantly on subjective, time-consuming manual gating. Machine learning approaches are being explored to support automated blast detection, yet the relative performance of anomaly detection versus supervised classification remains incompletely characterized.

**Methods.** We compared six computational approaches for blast identification: four anomaly detection algorithms (Isolation Forest, One-Class SVM, Local Outlier Factor, Half-Space Trees), a supervised Random Forest classifier, and an unsupervised UMAP-HDBSCAN clustering approach. The models were developed on the BLAST110 cohort (110 samples) using nested cross-validation with patient-level grouping, and externally validated on two independent LAIP29 samples spanning high (~41%) and low (~1%) blast prevalence.

**Results.** Random Forest achieved the strongest external validation performance (F1 = 0.965 diagnostic sample; F1 = 0.587 follow-up sample). Anomaly detection methods showed weaker, more variable performance, with substantial further decline at low blast prevalence (F1 as low as 0.0006); One-Class SVM performed best among anomaly detectors (F1 = 0.731, diagnostic sample). Several models showed a marked drop in performance between cross-validation and external validation. UMAP-HDBSCAN clustering failed to separate blast and non-blast populations into distinct clusters.

**Conclusions.** Supervised classification substantially outperformed the anomaly detection and unsupervised clustering approaches evaluated here, contrasting with prior work favoring anomaly detection for AML blast identification. This divergence may reflect differences in anomaly detection design; label-efficient approaches may still hold practical value where labeled data collection is limited.

**Key words.** Clinical flow cytometry, acute myeloid leukemia, anomaly detection, minimal residual disease, machine learning, blast identification

# Introduction

Flow cytometry is a cornerstone of the diagnostic and monitoring framework in the clinical management of acute myeloid leukemia (AML). By allowing single-cell analysis of light scatter and immunophenotypic profile to distinguish malignant cells from normal cells, it addresses the diagnosis, classification and subtyping, monitoring minimal residual disease (MRD), and relapse detection. In routine practice, however, clinicians still analyze this complex data mainly by manual gating on 2D plots (Figure 1). This approach is time-consuming, depends heavily on operator experience, and is difficult to scale to large datasets. It can also miss rare or subtle abnormal populations that are not obvious in standard biaxial views. In healthy individuals, blasts constitute ≤5% of nucleated cells in the bone marrow and are typically absent from the peripheral blood (Lichtman et al., 2022). AML is typically defined by the presence of ≥20% blasts in the bone marrow or peripheral blood (Döhner et al., 2022). However, according to the WHO 5$^{th}$ edition and the International Consensus Classification (ICC), AML can also be diagnosed with ≥10% blasts when AML-defining cytogenetic or molecular abnormalities are identified (Arber et al., 2022; Khoury et al., 2022). In these cases, the blast proportion can range from <5% to >20% depending on the diagnosis. In such datasets, blasts may appear as rare or outlier populations based on specific immunophenotypic markers, suggesting that anomaly or novelty detection models could be useful for identifying these abnormal cell populations.

Prior computational work in this space has explored automated approaches to flow cytometry analysis broadly, including cross-platform benchmarking of automated gating algorithms (Aghaeepour et al., 2013), probabilistic disease-progression assessment in AML (Rajwa et al., 2017), and clustering-based anomaly detection for cell-type identification (Pouyan et al., 2016). Within AML specifically, a UMAP-based anomaly detection approach has been proposed for minimal residual disease quantification, motivated by the argument that one-class classification methods may be better suited than supervised approaches given the scarcity and heterogeneity of AML training data (Weijler et al., 2022). A recent review has also surveyed the broader landscape of machine learning applications in clinical flow cytometry (Spies et al., 2025).

While supervised approaches are generally preferred when labelled data are available, there has been limited work applying anomaly detection methods to FCS-based leukemia datasets and directly comparing them with supervised and clustering-based approaches. Notably, the dataset used in this study comes from Mocking et al. (2024). Their study compared novelty detection, a method that flags cells that don't match a known "normal" pattern, against UMAP-HDBSCAN clustering. They tested this using a Gaussian mixture reference model, local outlier factor, and one-class SVM. Separately, they benchmarked several supervised classifiers for blast identification. Their novelty detection method worked by first learning what healthy, non-leukemic cells typically look like, using a separate set of samples from disease-free patients. Any cell that did not match this healthy pattern was then flagged as potentially leukemic. Supervised classification was not evaluated within the same benchmark as this novelty detection approach.

Here, we evaluate anomaly detection, supervised classification, and unsupervised clustering together on a single blast identification task, without requiring a dedicated disease-free reference cohort. Instead, our anomaly detection models learn what "normal" looks like directly from the same mixed patient samples being classified, treating rare or unusual cells within that sample as potential blasts, rather than comparing against a separate healthy reference group. We find that supervised classification substantially outperforms both anomaly detection and unsupervised clustering across the range of blast prevalence relevant to diagnosis and MRD monitoring. This result contrasts with prior work favoring anomaly detection for this setting, and we discuss how this difference in anomaly detection design, using patient samples themselves versus a separate healthy reference, may help explain this divergence.

## Methods

**Data source.**

The dataset was obtained from the Zenodo public repository (https://zenodo.org/records/11046402), comprising two cohorts as described by Mocking et al. (2024): BLAST110, gated to identify immature myeloid blasts (110 samples: 20 healthy controls, 30 AML diagnosis samples, and 60 follow-up samples spanning MRD-low, -intermediate, and -high categories, approximately 20 each); and LAIP29, gated to identify leukemia-associated immunophenotype (LAIP+) populations (48 samples: 28 diagnosis and 20 follow-up). Each sample was measured across four antibody panels (P1–P4), each comprising eight markers (five common "backbone" markers plus three panel-specific markers). Of the expected 440 BLAST110 files (four panels per patient), only 403 were available; similarly, of the expected 160 LAIP29 files, 83 were available. The difference in the expected vs. available is due to the sample files missing.

**Data Preprocessing.**

The preprocessing followed a four-step workflow that was applied uniformly across all samples. In the first step of **Margin (event) removal**, the events at or beyond the detector's dynamic range (arising from detector saturation, electronic noise, or fluidic disturbances) were removed using PeacoQC::RemoveMargins in R, applied to the primary scatter channels (FSC-A, FSC-H, SSC-A, SSC-H). In the second step of **Debris removal**, the non-cellular or fragmented material, typically exhibiting low forward and side scatter, was removed using a two-step approach: the first is automated detection via the nmRemove() function (flowDensity package), followed by the quantile-based filtering excluding the lowest 1% of events in both FSC-A and SSC-A. The third step involved **Doublet removal**. Here, the events representing multiple co-detected cells were excluded using PeacoQC's RemoveDoublets() function, based on atypical scatter/fluorescence ratios.

Logicle transformation is a way of rescaling fluorescence intensity values so that both very low and very high signals can be displayed and analyzed clearly, avoiding the distortion that occurs when standard logarithmic scaling is applied to near-zero or negative values (Parks et al., 2006). Finally, in the **Transformation and scaling** step, the logicle transformation was applied using the estimateLogicle function (flowCore), with transformation parameters estimated per channel and aggregated by median to produce a tube-specific transformList, consistent with the transformation strategy of Mocking et al. (2024). Scatter parameters (FSC-A, FSC-H, SSC-A, SSC-H) subsequently underwent min-max scaling to a [0,1] range based on the 1st and 99th percentiles.

**Downsampling**

To standardize the dataset size and manage computational load, a uniform random downsampling approach was performed independently for each FCS file. Downsampling schemes varied by model: Isolation Forest (downsampled to 5,000 events/file, then further adjusted to maintain a 5–7% blast proportion), One-Class SVM (2,000 events/file), Random Forest (5,000 events/file), Half-Space Trees (5,000 events/file, resampled to a 5% blast proportion), and Local Outlier Factor (2,000–5,000 events/file, trained on healthy reference samples).

**Feature Selection**

Feature selection was restricted to backbone parameters consistently available across all the four antibody panels, ensuring cross-sample comparability (Table 1). The four markers (SSC-A, CD45, CD34, CD117) were retained for their established biological relevance in blast gating. CD13 was added as a fifth marker given its role as a myeloid lineage marker strongly associated with AML. HLA-DR, although also a backbone marker, was excluded, as it primarily reflects progenitor or stem cell status rather than leukemia-specific discrimination. Kernel density estimation (KDE) analysis of marker distributions supported this selection, where CD34 and CD117 showed the strongest separation between blast and non-blast populations, CD13 and CD45 showed good but incomplete separation, and FSC-A showed poor discriminatory power (Figures 2 and 3). This five-marker set (SSC-A, CD34, CD45, CD117, CD13) was used for all the models, with the exception of UMAP-HDBSCAN clustering, which used a four-marker subset excluding CD13 (see Limitations).

**General Model Framework**

All the models were validated using a 5-fold nested cross-validation framework to obtain an unbiased estimate of generalization performance on the BLAST110 cohort. Using a nested cross-validation technique, the outer cross-validation loop assessed the performance on held-out data, while the inner cross-validation loop optimized hyperparameters using the corresponding training data. The patient identifiers were used as the grouping variable in both the loops to ensure that all the samples from a given patient were confined to a single fold, preventing information leakage between the training and the evaluation sets.

For Isolation Forest, Local Outlier Factor (LOF), Half-Space Trees (HST), and Random Forest, hyperparameter optimization was conducted via a grid search within the inner loop, and the final hyperparameters were determined by consensus selection across the outer folds. For Isolation Forest, HST, and Random Forest, this meant the hyperparameter configuration selected most frequently across the five folds was adopted as final. For LOF, whose hyperparameter grid rarely produced exactly repeated configurations across folds, consensus was instead determined by taking the most common value for each hyperparameter (n_neighbors and contamination) independently and combining them into the final parameter set. One-Class SVM followed a different approach, where given the one-class training setup, hyperparameters were tuned without a grid search, with the configuration selected based on the highest mean Area Under the Receiver Operating Characteristic Curve (AUROC) across the folds.

For each outer fold, the classification threshold was optimized by maximizing the F-measure. The final decision threshold was defined as the median of fold-wise optimal thresholds for all models except Random Forest, which used the mean.

External validation was performed on two independent samples from the LAIP29 cohort: a diagnostic sample (~41% blasts) and a follow-up sample (~1% blasts). Model interpretability was evaluated using SHAP analysis on randomly sampled events from the diagnostic dataset, and Cohen's kappa was computed to quantify agreement between model predictions and the manually gated reference labels provided with the Mocking et al dataset.

**Model-Specific Details**

**Random Forest.** Random Forest (Breiman, 2001)is a supervised ensemble classifier that aggregates predictions from many decision trees, each trained on a bootstrap sample of the data with a random subset of features considered at each split. Unlike the anomaly detection methods, Random Forest was trained directly on labeled blast/non-blast events. Hyperparameters, including the number of estimators, maximum tree depth, minimum samples per split, and minimum samples per leaf, were tuned via grid search within the inner cross-validation loop, using average precision as the optimization metric. Feature importance was assessed using both Gini-based importance and SHAP analysis.

**One-Class SVM**. One-Class SVM (OC-SVM) (Schölkopf et al., 2001)is an anomaly detection method trained exclusively on non-blast events, learning a boundary that encloses the normal population; events falling outside this boundary at test time are flagged as blasts. The model used a radial basis function (RBF) kernel, with the kernel bandwidth (gamma) and the ν parameter, which controls the expected proportion of outliers, tuned within the inner cross-validation loop using AUROC as the selection metric, since grid search was not applicable given the one-class training setup.

**Isolation Forest.** Isolation Forest (Liu et al., 2008)is an anomaly detection method that isolates observations through recursive random partitioning; anomalous events, being distinct from the majority, typically require fewer splits to isolate and are assigned higher anomaly scores. The model was trained on data maintaining a low blast proportion (approximately 5–7% of events) to preserve the rare-event detection setting. Hyperparameters, including the number of trees, tree dimensionality, sample size, and contamination range, were tuned within the inner cross-validation loop by maximizing the area under the precision-recall curve (AUPRC), given the class imbalance inherent to rare-event detection.

**Local Outlier Factor (LOF)**. Local Outlier Factor (Breunig et al., 2000)identifies outliers by comparing the local density around a point to the local density of its neighbors; points in sparser regions relative to their neighbors are assigned higher LOF scores and flagged as anomalous. Hyperparameters, specifically the number of neighbors (n_neighbors) and contamination level, were tuned within the inner cross-validation loop via grid search, with the classification threshold optimized by maximizing the F1-measure.

**Half-Space Trees (HST).** Half-Space Trees (Tan et al., 2011)is a one-class, tree-based anomaly detection method that recursively partitions the feature space using random axis-aligned splits; anomalies are identified in shallow, low-mass regions of the resulting trees. The model was trained exclusively on non-blast events from a downsampled dataset with a 5% blast proportion. Hyperparameters, including the number of trees and tree depth, were tuned within the inner cross-validation loop by maximizing the F1-measure after anomaly score threshold tuning.

**UMAP-HDBSCAN.** UMAP (Uniform Manifold Approximation and Projection) (McInnes et al., 2018)is a manifold learning technique that projects high-dimensional data into a lower-dimensional space while preserving local neighborhood structure; HDBSCAN (Hierarchical Density-Based Spatial Clustering of Applications with Noise) (McInnes et al., 2017)then clusters this embedding by identifying regions of high density, labeling sparse regions as noise, without requiring the number of clusters to be specified in advance. Unlike the anomaly detection and supervised models, UMAP-HDBSCAN used a four-marker subset (SSC-A, CD45, CD34, CD117), excluding CD13 (see Limitations). UMAP was configured with three output dimensions, 30 neighbors, and a minimum distance of 0, followed by HDBSCAN clustering with a minimum cluster size of 50. Clusters were evaluated within a 5-fold cross-validation scheme with patient-level grouping, and cluster identity was compared against manually gated blast labels to assess the correspondence between cluster membership and blast status.

## Results

Identifying leukemic blasts accurately across a wide range of prevalence, from diagnosis (where blasts are abundant) to MRD monitoring (where they are rare), is central to clinical AML management. We therefore evaluated six computational approaches spanning three methodological families, anomaly detection (Isolation Forest, One-Class SVM, Local Outlier Factor, Half-Space Trees), supervised classification (Random Forest), and unsupervised clustering (UMAP-HDBSCAN), using nested cross-validation on the BLAST110 cohort and external validation on two independent LAIP29 samples representing high (~41%) and low (~1%) blast prevalence.

### Random Forest achieves the strongest and most robust performance across cross-validation and external validation

In the nested cross-validation, Random Forest achieved the highest performance among all models evaluated (Table 2), with a mean accuracy of 0.970 (± 0.007), F1-score of 0.911 (± 0.019), and AUROC of 0.994 (± 0.002). In the external validation, the performance remained strong on the diagnostic sample (~41% blasts, Table 3a), with an F1-score of 0.965 and AUROC of 0.994. On the follow-up sample (~1% blasts, Table 3b), F1 declined to 0.587, reflecting the difficulty of maintaining precision-recall balance at low blast prevalence; AUROC remained high at 0.998, indicating the model retained strong overall ranking ability across decision thresholds, a distinction from F1's dependence on a single fixed threshold.

### Anomaly detection models generalize inconsistently, collapsing at low blast prevalence

In the cross-validation, anomaly detection models showed a wide range of performance, with F1-scores ranging from 0.229 (LOF) to 0.718 (One-Class SVM), and Isolation Forest and HST falling in between (F1 = 0.580 and 0.459, respectively). Generalization to external validation varied markedly by model and by blast prevalence. In the diagnostic sample (~41% blasts), One-Class SVM maintained performance comparable to cross-validation (F1 = 0.731), while Isolation Forest and HST declined substantially (F1 = 0.088 and 0.354, respectively), and LOF failed to detect any blasts at all (F1 = 0). In the low-prevalence follow-up sample (~1% blasts), performance declined sharply across all four models, including One-Class SVM (F1 = 0.194, 0.084, 0.0006, and 0.141 for Isolation Forest, One-Class SVM, LOF, and HST, respectively), indicating that even the comparatively strongest anomaly detector was unable to maintain reliable precision-recall balance once blast prevalence dropped to clinically realistic MRD levels.

**Among anomaly detection methods, One-Class SVM achieves the best precision-recall trade-off** Among the four anomaly detection models, One-Class SVM achieved the highest cross-validation F1-score (0.718), driven by a recall of 0.911 and precision of 0.594. This represents the most favorable balance between sensitivity and precision among the anomaly detectors evaluated, though precision remained substantially lower than recall, reflecting a tendency toward over-flagging non-blast events as anomalous even in its best-performing configuration. By comparison, Isolation Forest (F1 = 0.580), HST (F1 = 0.459), and LOF (F1 = 0.229) each showed weaker balance between precision and recall in cross-validation.

**Unsupervised clustering (UMAP-HDBSCAN) fails to separate blast and non-blast populations**

Across all the five cross-validation folds, the UMAP embedding formed a continuous, smooth manifold without clearly separated regions (Figure 4), with manually labelled blast events broadly distributed across the embedding rather than confined to distinct areas. When clustered using HDBSCAN, almost all events, both blast and non-blast, were assigned to a single dominant high-density cluster, with only a few small secondary clusters and noise points. Consequently, the cluster membership did not correspond to blast status in any fold, indicating that purely unsupervised density-based clustering on this feature set is unable to recover blast populations as distinct clusters.

**CD117 and CD34 emerge as the dominant contributors to blast classification**

Feature importance analysis of the final Random Forest model identified CD117 (PC7-A) and CD34 (PerCP-A) as the dominant contributors to blast classification (Figure 5), followed by CD45 and SSC-A, with CD13 contributing minimally. This pattern was consistent across both the BLAST110 and LAIP29 cohorts. SHAP analysis further indicated that elevated CD117 and CD34 expression was strongly associated with blast classification (Figure 6), while CD45 showed a non-linear relationship, with lower or intermediate expression predictive of blast status, consistent with the characteristically dim CD45 expression observed in leukemic blasts.

# Discussion

Manual gating remains the standard approach for identifying leukemic blasts in flow cytometry data, but it is time-consuming, subjective, and difficult to scale, particularly as blast populations become rare in the context of MRD monitoring. This motivated our evaluation of anomaly detection as an alternative to supervised classification, based on the premise that leukemic blasts, being immunophenotypically distinct and often present in low proportions, might behave as detectable outliers relative to the surrounding non-blast population. We compared four anomaly detection algorithms, a supervised Random Forest classifier, and an unsupervised UMAP-HDBSCAN clustering approach across cross-validation and external validation settings spanning a wide range of blast prevalence.

Random Forest substantially outperformed both the anomaly detection algorithms and the unsupervised clustering approach across every evaluation setting, achieving near-perfect discrimination in cross-validation (F1 = 0.911, AUROC = 0.994) and maintaining strong performance on external validation, even at low blast prevalence (F1 = 0.587 on the ~1% blast follow-up sample). By contrast, the anomaly detection methods showed substantially weaker and less consistent performance, with F1-scores at low prevalence ranging from 0.0006 to 0.194, and the unsupervised UMAP-HDBSCAN approach failed to separate blast and non-blast populations into distinct clusters altogether. Taken together, these results indicate that, at least for the panel and feature set evaluated here, supervised classification substantially outperforms both anomaly detection and unsupervised clustering for blast identification across the two blast prevalence regimes evaluated here, from diagnostic-level to MRD-level blast burden.

The comparatively weaker performance of anomaly detection methods can be understood in light of the underlying feature distributions. Kernel density estimation analysis of the five backbone markers revealed that no single marker, nor the full combination of markers considered jointly, achieved clean separation between blast and non-blast populations: event-level propagation analysis found that no cells remained exclusively separated across all features simultaneously, with a final blast retention rate of 0 % once all markers were considered together. This substantial feature overlap likely undermines the core assumption of anomaly detection, namely that blasts occupy a distinct, separable region of feature space relative to the non-blast population. Where this assumption does not hold, methods that rely on modeling the boundary of a "normal" population, or on isolating rare observations through recursive partitioning, are likely to misclassify blasts that fall within regions of substantial overlap with non-blast cells, regardless of how well the model's hyperparameters are tuned.

These findings partially converge with, and partially diverge from, observations in the study that originally generated the BLAST110 and LAIP29 datasets. Mocking et al. (2024) reported that a supervised classifier (a Gaussian mixture model-based classifier, GMMclf) outperformed other supervised methods for general blast identification, consistent with the strong performance of supervised classification observed in our study. For identifying leukemic (LAIP+) blasts specifically, their GMM-based novelty detection outperformed both local outlier factor and one-class SVM, as well as UMAP-HDBSCAN clustering; notably, their UMAP-HDBSCAN approach also failed to recover leukemic populations as distinct clusters in a substantial proportion of samples (34–44%, depending on setting), converging with our own UMAP-HDBSCAN findings. However, their overall framing of novelty detection as a viable and favorable alternative to supervised classification for AML's heterogeneous phenotypes contrasts with the weaker anomaly detection performance observed in our study, a tension explored further below.

This finding also stands in contrast with Weijler et al. (2022), who argued that one-class classification approaches may be better suited than supervised methods for AML MRD quantification. Their reasoning was based on AML's heterogeneity reflected in the more than 100 distinct leukemia-associated immunophenotypes described, some observed in fewer than 1% of patients (Mocking et al., 2024). Our results, evaluated on a single dataset with a fixed feature set and a relatively narrow range of blast phenotypes, do not support this expectation. This discrepancy may reflect differences in study design, specifically our anomaly detection models defined outlier status based on within-sample rarity rather than deviation from a dedicated healthy reference population. Whereas, Weijler et al.'s approach relied on a reference-based framework, rather than a fundamental limitation of anomaly detection as a broader strategy. Reconciling these findings, particularly whether reference-based versus within-sample anomaly detection designs perform differently as phenotypic heterogeneity increases, represents an important direction for future work.

The prominence of CD117 and CD34 in Random Forest's feature importance and SHAP analyses is consistent with established AML blast immunophenotype. Both markers are characteristically expressed on immature myeloid blasts, CD34 marking early hematopoietic progenitors and CD117 marking stem cell factor receptor expression, and their combined presence, alongside dim CD45 expression, is a recognized hallmark of the immature immunophenotypic profile used in manual blast gating (Zeijlemaker et al., 2019). The model's reliance on these markers therefore aligns with established biological reasoning for blast identification, suggesting that the model's decisions are grounded in biologically interpretable signal, rather than spurious patterns in the data.

Random Forest's strong performance in this study depends on the availability of labeled training data, a requirement that carries its own practical limitations. As noted by Mocking et al. (2024), supervised models trained on a specific antibody panel and instrument platform may not generalize well to new centers or platforms without recalibration, given the substantial phenotypic heterogeneity across AML patients. Anomaly detection's original appeal, label-efficient training requiring only unlabeled or partially labeled data, therefore remains a meaningful practical consideration, though even the anomaly detection approaches evaluated here required some labeled data for hyperparameter tuning and threshold selection, reducing but not eliminating this advantage in practice. These findings may be particularly relevant to healthcare settings with decentralized medical record systems, heterogeneous flow cytometry infrastructure, or limited capacity to build large, center-specific labeled datasets, where the label efficiency originally motivating interest in anomaly detection remains an important practical consideration despite its weaker performance in this study.

This study has several limitations. It is a proof-of-concept analysis based on a single publicly available dataset, and the generalizability of these findings to other AML cohorts, flow cytometry platforms, or institutions remains to be established. The feature set was restricted to five backbone markers common across all panels; UMAP-HDBSCAN, in particular, was evaluated using a four-marker subset that excluded CD13, a difference from the other models that should be considered when interpreting its results relative to the other methods, alongside the feature overlap discussed above as the primary driver of its failure.

Unlike Mocking et al. (2024), this study did not employ a dedicated disease-free reference cohort for novelty detection; instead, anomaly detection models defined outlier status based on within-sample rarity in mixed clinical samples, a design difference that may partly explain the divergence from prior findings.

Additionally, none of the models evaluated here were tested for robustness to variation in marker panels or fluorophore conjugation across laboratories: an additional or different set of non-backbone markers alongside the same core panel would likely have minimal impact, since the models rely only on the backbone features, but differences in fluorophore conjugation of the same markers could meaningfully shift measured intensity distributions and may require recalibration or retraining before these models could be applied at a different center. Finally, downsampling strategies varied across models, and while patient-level grouping was maintained throughout to prevent data leakage, differences in per-model data volume may have contributed to some of the performance variation observed.

Future work should prioritize validation of these findings on independent AML cohorts and flow cytometry platforms to establish generalizability beyond the single dataset evaluated here. Given the divergence from Weijler et al. (2022) discussed above, a direct comparison of reference-based and within-sample anomaly detection designs on the same dataset would help clarify whether the choice of reference strategy, rather than anomaly detection as a broader approach, accounts for the difference in findings. Testing model robustness to variation in marker panels and fluorophore conjugation, particularly whether retraining or recalibration is required across sites, would also help establish practical deployment feasibility. Finally, hybrid approaches that combine the label efficiency of anomaly detection with the discriminative strength of supervised classification, for example, using anomaly detection for initial screening followed by supervised refinement, may offer a promising direction for settings where fully labeled training data remains difficult to obtain.

**This manuscript is currently under review, all authors have seen and approved this manuscript, and it hasn't been accepted or published elsewhere.**

## Data and code availability

The flow cytometry datasets used in this study (BLAST110 and LAIP29) are publicly available. No new patient data were generated for this work. All analysis scripts, including data preprocessing, model training, cross-validation, and external validation pipelines, are openly available at https://github.com/raghuyennamalli/flow_cyto_AML

## Acknowledgements

RMY acknowledges Jawaharlal Nehru University for infrastructural support. RSA acknowledges the Ashoka University for infrastructural support and gratefully acknowledges the Ashoka University HPC facility for providing computational resources used in this work.

## Funding

Our gratitude extends to Mphasis AI & Applied Tech Lab at Ashoka supported by the Mphasis F1 Foundation for their generous support and funding.

## Conflict of interest disclosure

All authors declare that they have no financial or non-financial conflicts of interest related to this manuscript.

**Table 1. Antibody panel configuration across the four flow cytometry tubes (P1–P4)**

| **Channel** | **P1** | **P2** | **P3** | **P4** |
|---|---|---|---|---|
| FITC | CD7 | CD15 | CD36 | CD2 |
| PE | CD56 | CD22 | CD14 | CD133 |
| PerCP-Cy5.5 | **CD34*** | **CD34*** | **CD34*** | **CD34*** |
| PC7 | **CD117*** | **CD117*** | **CD117*** | **CD117*** |
| APC | CD33 | CD19 | CD11b | CD33 |
| APC-H7 | *HLA-DR†** | *HLA-DR†** | *HLA-DR†** | *HLA-DR†** |
| BV421 | **CD13*** | **CD13*** | **CD13*** | **CD13*** |
| KO | **CD45*** | **CD45*** | **CD45*** | **CD45*** |

***** Backbone marker (present identically across all four panels).

**†** Excluded from the final analytical feature set; HLA-DR primarily reflects general progenitor/stem cell status rather than leukemia-specific signal, and was not retained despite being a backbone marker (see Methods, Feature selection).

Non-daggered, non-bolded entries are panel-specific markers, unique to each tube.

**Table 2. Cross-validation performance (BLAST110 cohort)**

| Metric | Isolation Forest | One-Class SVM | LOF | HST | Random Forest |
|---|---|---|---|---|---|
| Trained on | Blast & non-blast events | Non-blast events only | Blast & non-blast events | Non-blast events only | Blast & non-blast events |
| Training blast prevalence | 5.63% | 16.61% | 15.19% | 5.70% | 16.64% |
| Accuracy | 0.945 ± 0.012 | 0.871 ± 0.029 | 0.770 ± 0.025 | 0.926 ± 0.0188 | 0.970 ± 0.007 |
| Precision | 0.507 ± 0.059 | 0.594 ± 0.085 | 0.233 ± 0.050 | 0.389 ± 0.048 | 0.878 ± 0.050 |
| Recall | 0.679 ± 0.060 | 0.911 ± 0.035 | 0.227 ± 0.036 | 0.575 ± 0.157 | 0.950 ± 0.022 |
| F1-score | 0.580 ± 0.054 | 0.718 ± 0.042 | 0.229 ± 0.038 | 0.459 ± 0.073 | 0.911 ± 0.019 |
| AUROC | 0.955 ± 0.007 | 0.938 ± 0.000 | 0.593 ± 0.019 | 0.940 ± 0.013 | 0.994 ± 0.002 |
| AUPRC | 0.506 ± 0.073 | 0.699 ± 0.001 | 0.200 ± 0.040 | 0.3556 ± 0.045 | 0.967 ± 0.010 |
| Cohen's kappa | 0.551 ± 0.049 | 0.305 ± 0.158 | 0.0036 ± 0.007 | 0.421 ± 0.080 | 0.893 ± 0.019 |

**Table 3 - External validation performance (LAIP29 cohort)**

**3a. Diagnostic sample (~41% blasts)**

| Metric | Isolation Forest | One-Class SVM | LOF | HST | Random Forest |
|---|---|---|---|---|---|
| Accuracy | 0.584 | 0.777 | 0.588 | 0.633 | 0.971 |
| Precision | 0.455 | 0.728 | 0 | 0.646 | 0.973 |
| Recall | 0.049 | 0.733 | 0 | 0.244 | 0.957 |
| F1-score | 0.088 | 0.731 | 0 | 0.354 | 0.965 |
| Specificity | 0.959 | 0.77 | 1.00 | 0.906 | 0.982 |
| AUROC | 0.207 | 0.162 | 0.420 | 0.906 | 0.994 |
| AUPRC | 0.280 | 0.27 | 0.353 | 0.851 | 0.980 |
| Cohen's kappa | 0.009 | 0.541 | 0.00004 | 0.166 | 0.941 |

**3b. Follow-up sample (~1% blasts)**

| Metric | Isolation Forest | One-Class SVM | LOF | HST | Random Forest |
|---|---|---|---|---|---|
| Accuracy | 0.929 | 0.772 | 0.989 | 0.873 | 0.985 |
| Precision | 0.110 | 0.044 | 0.006 | 0.076 | 0.417 |
| Recall | 0.818 | 1.000 | 0.0003 | 0.998 | 0.993 |
| F1-score | 0.194 | 0.084 | 0.001 | 0.141 | 0.587 |
| Specificity | 0.930 | 0.749 | 0.999 | 0.871 | 0.985 |
| AUROC | 0.039 | 0.026 | 0.828 | 0.871 | 0.998 |
| AUPRC | 0.010 | 0.010 | 0.051 | 0.226 | 0.730 |
| Cohen's kappa | 0.178 | 0.065 | 0.0004 | 0.124 | 0.581 |

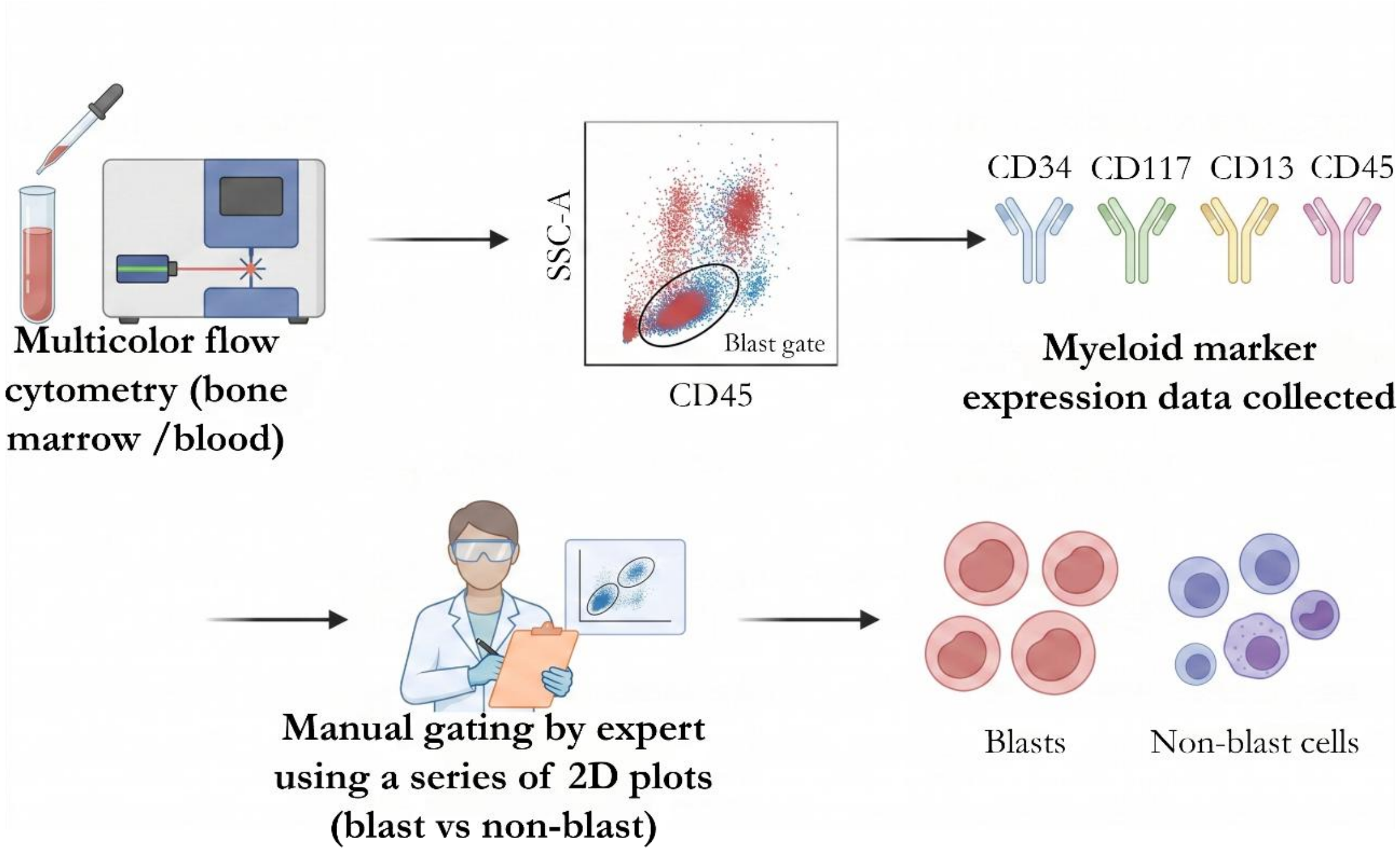

**Figure 1. Manual gating workflow for identifying leukemic blasts in flow cytometry data.** An initial gate on CD45 versus side scatter (SSC) identifies a candidate blast population, refined using CD34, CD117, and CD13 before expert classification into blast versus non-blast events, consistent with the feature set used in this study's computational analyses (see Methods, Feature Selection).

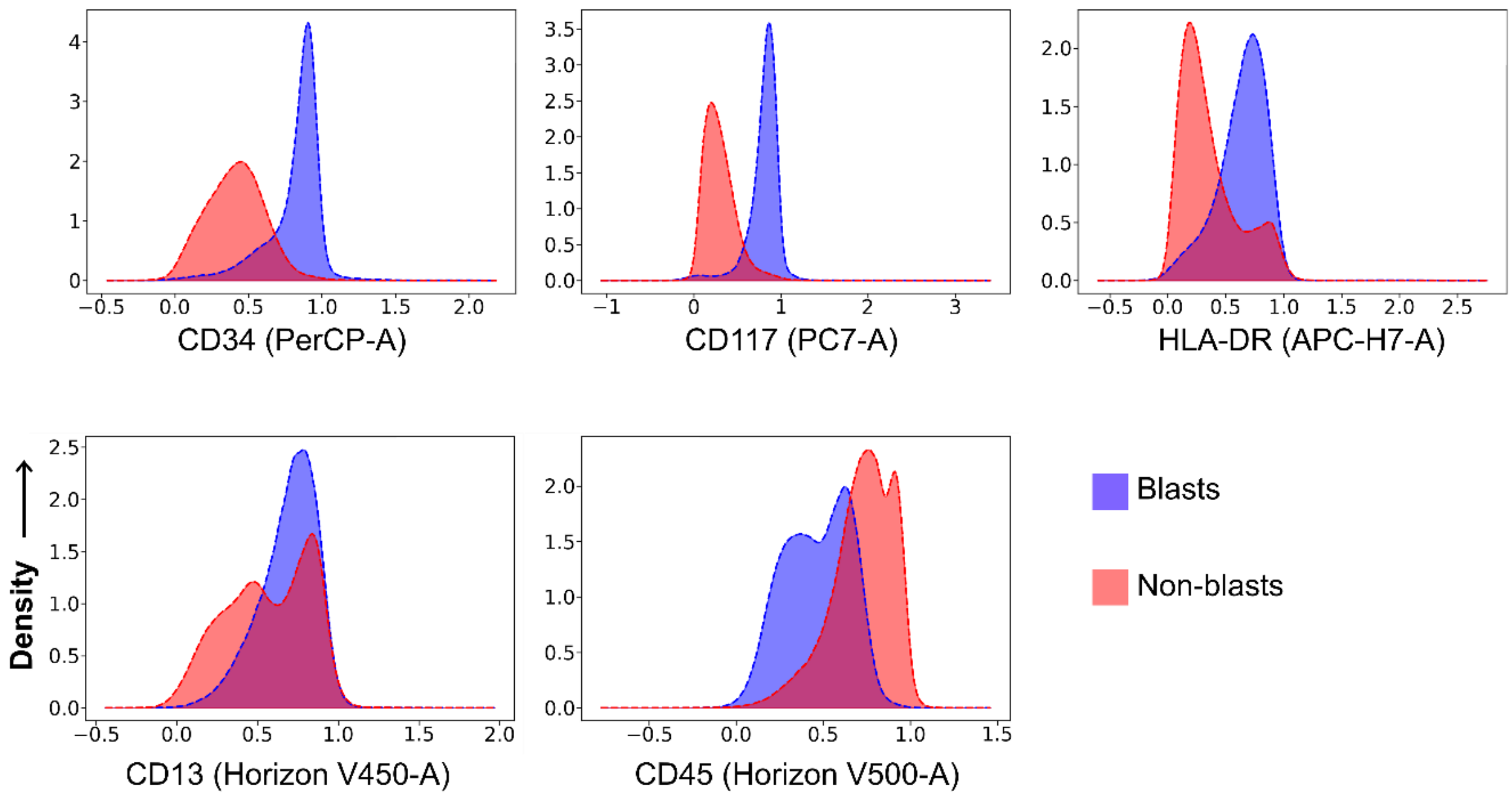

**Figure 2. Kernel density estimation (KDE) of fluorescence intensity for blast (blue) versus non-blast (red) events across five immunophenotypic markers (CD34, CD117, HLA-DR, CD13, CD45).** Each panel shows smoothed probability density curves, normalized separately within each population such that the area under each curve integrates to 1; curve height therefore reflects the relative concentration of events at a given intensity, not absolute event counts. Purple shading indicates regions of overlap between the two distributions.

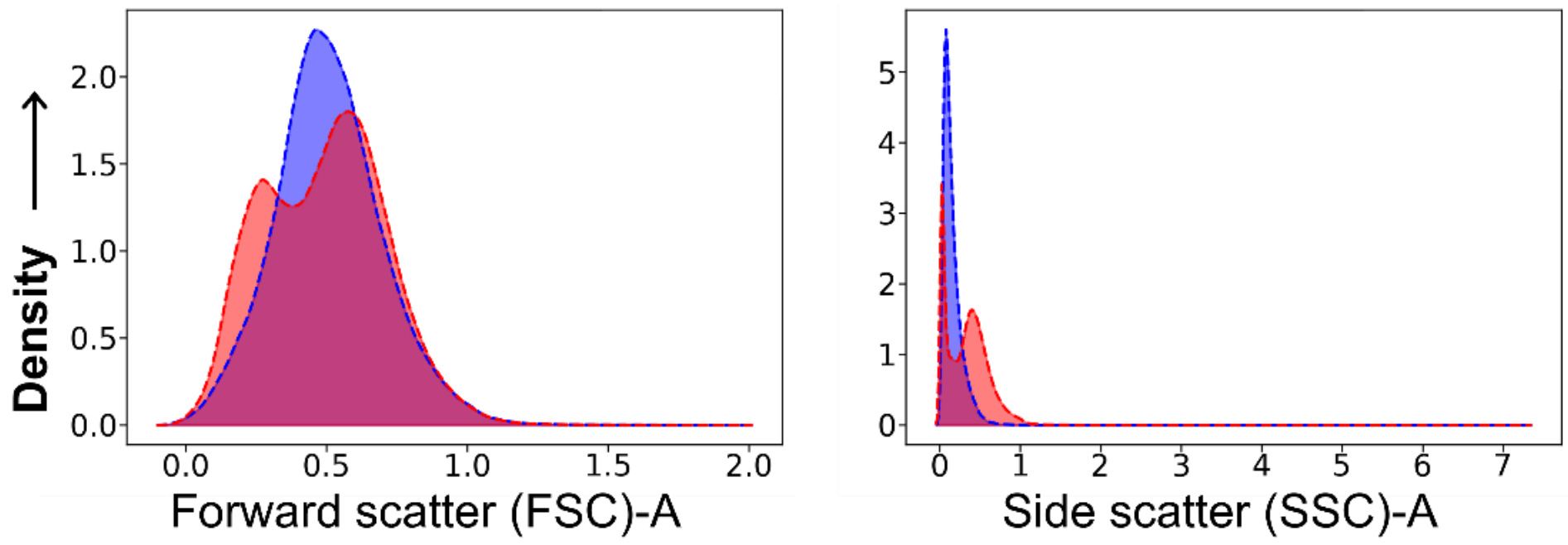

**Figure 3. Kernel density estimation (KDE) of forward and side scatter for blast (blue) versus non-blast (red) events (FSC-A, SSC-A).** Each panel shows smoothed probability density curves, normalized separately within each population such that the area under each curve integrates to 1; curve height therefore reflects the relative concentration of events at a given intensity, not absolute event counts. Purple shading indicates regions of overlap between the two distributions.

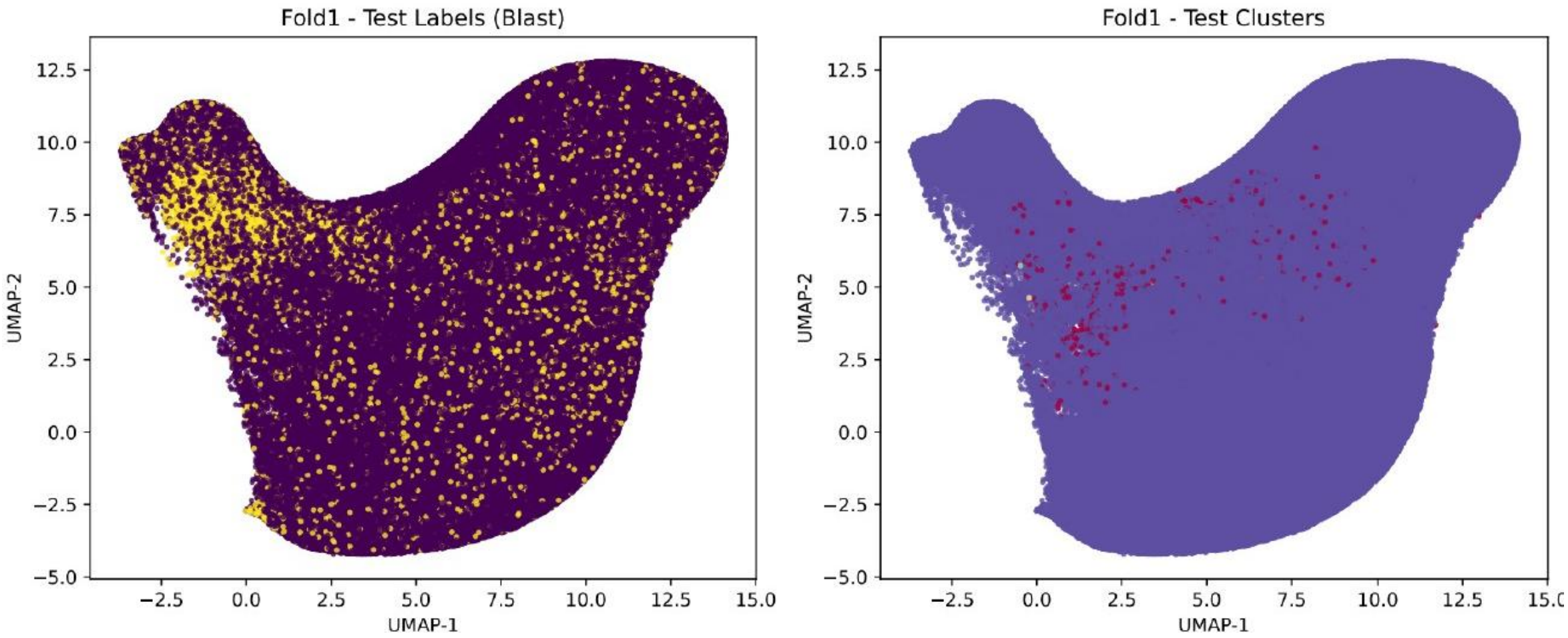

**Figure 4. UMAP embedding of test events (fold 1).** Left: events colored by manual blast label (yellow = blast, purple = non-blast), showing substantial overlap between the two populations. Right: events colored by HDBSCAN cluster assignment (red = dominant dense cluster, purple = background), indicating that unsupervised clustering does not separate blasts into a distinct group.

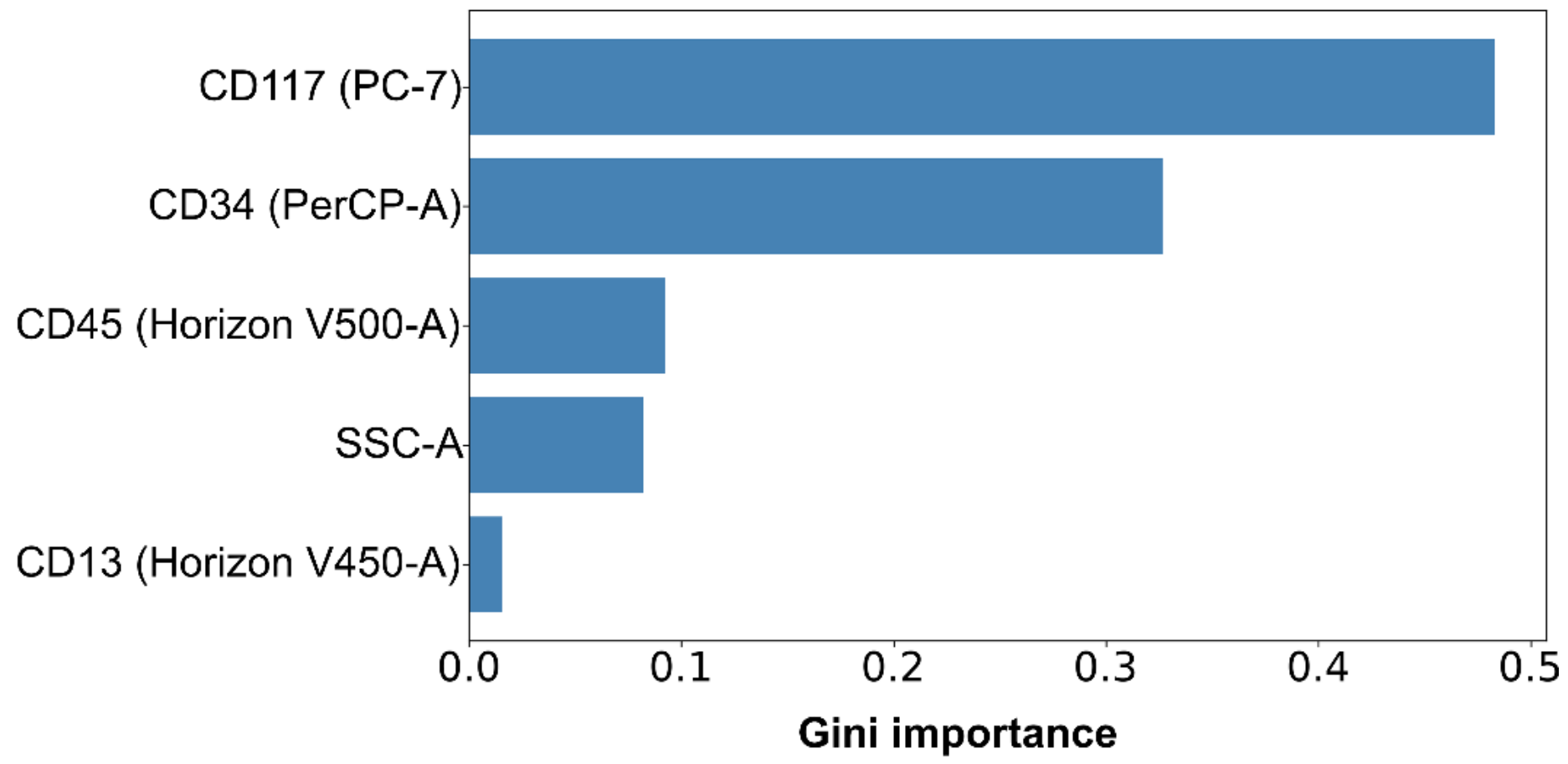

**Figure 5. Feature importance in the final Random Forest model, BLAST110 cohort.** PC7-A (CD117) and PerCP-A (CD34) contribute most strongly to blast classification, followed by Horizon V500-A (CD45) and SSC-A, with Horizon V450-A (CD13) contributing minimally.

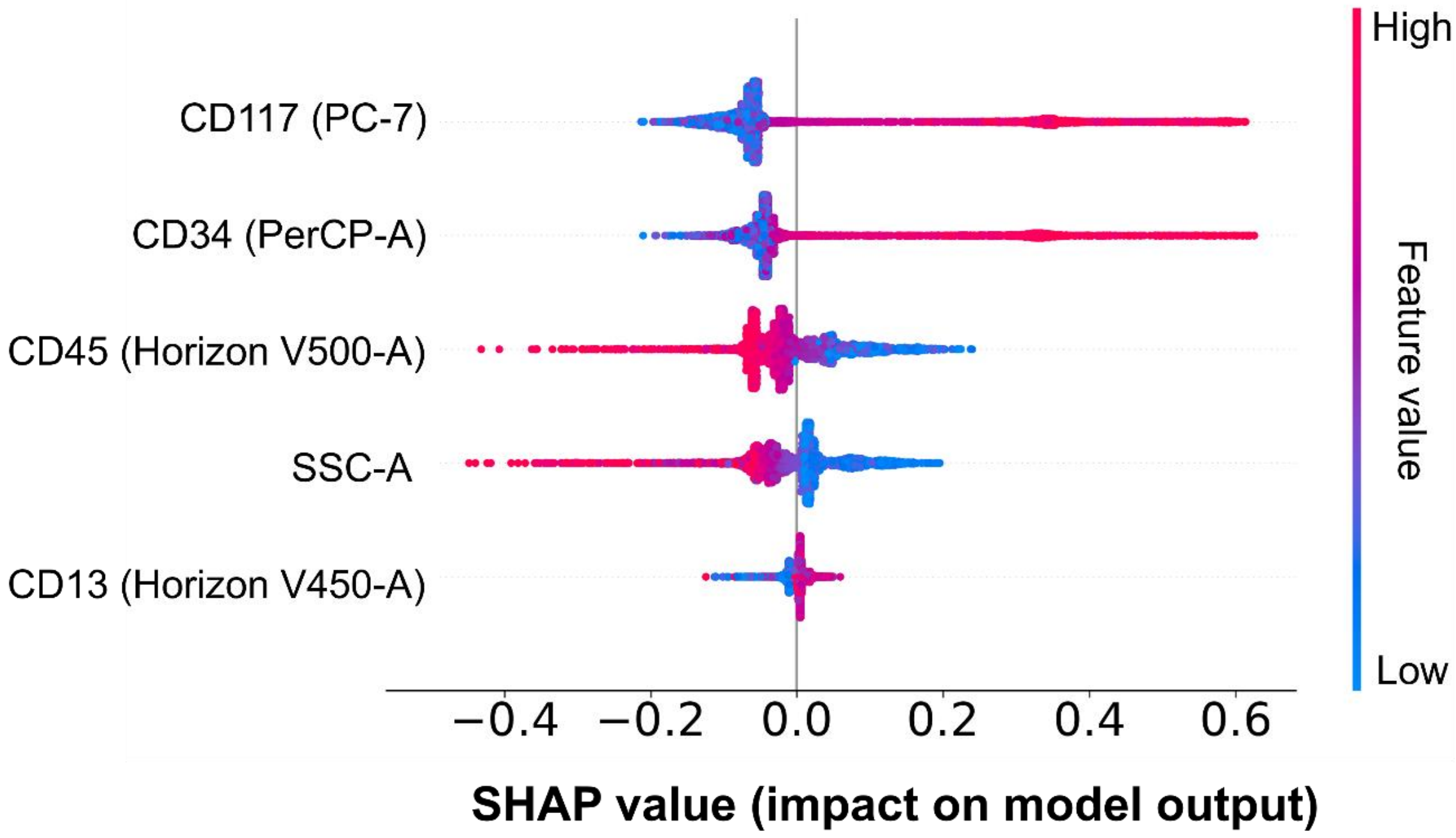

**Figure 6. SHAP summary plot showing the impact of each feature on Random Forest blast predictions, BLAST110 cohort.** Each point represents one cell; position along the x-axis indicates the magnitude and direction of that feature's contribution to the model's prediction for that cell, and color indicates the feature's expression level (red = high, blue = low). High CD117 (PC7-A) and CD34 (PerCP-A) expression strongly increases predicted blast probability. CD45 (Horizon V500-A) shows a non-linear relationship, with lower/intermediate expression increasing predicted blast probability.